\documentclass[aps, prd, amsmath, floats, floatfix, twocolumn, %
superscriptaddress, nofootinbib, showpacs]{revtex4}

\usepackage{graphicx}
\usepackage[usenames,dvipsnames]{color}
\usepackage{bm}
\usepackage[colorlinks=true, plainpages=false]{hyperref}
\definecolor{CiteColor}{rgb}{0, 0.75, 0}
\hypersetup{citecolor=CiteColor} %
\definecolor{RefColor}{rgb}{0.75, 0, 0}
\hypersetup{linkcolor=RefColor} %

\newcommand{\Eadm}{M_{\text{ADM}}}

\newcommand{\ra}{\ensuremath{r_{\text{areal}}}}

\newcommand{\gammaHtwo}{\gamma^{{}_H}_2}

\newcommand{\Caltech}{\affiliation{Theoretical Astrophysics 130-33,
    California Institute of Technology, Pasadena, California 91125}}
\newcommand{\Cornell}{\affiliation{Center for Radiophysics and Space
    Research, Cornell University, Ithaca, New York 14853}}

\begin{document}
\vspace{-2.5cm} 

\title{High-accuracy waveforms for
       binary black hole inspiral, merger, and ringdown}

\author{Mark A. Scheel} \Caltech
\author{Michael Boyle} \Caltech
\author{Tony Chu} \Caltech
\author{Lawrence E. Kidder} \Cornell
\author{Keith D. Matthews} \Caltech
\author{Harald P. Pfeiffer} \Caltech

\date{\today}

\begin{abstract}
  The first spectral numerical simulations of 16 orbits, merger, and
  ringdown of an equal-mass nonspinning binary black hole system are presented.
  Gravitational waveforms from these simulations have accumulated
  numerical phase errors through ringdown of $\lesssim 0.1$
  radian when measured from the beginning of the simulation, and
  $\lesssim 0.02$ radian when waveforms are time and phase shifted to
  agree at the peak amplitude. The waveform seen by an observer
  at infinity is determined from waveforms computed at
  finite radii by an extrapolation process 
  accurate to $\lesssim 0.01$ radian in phase.  
  The phase difference between this waveform
  at infinity and the waveform measured at a finite radius of $r=100M$ is
  about half a radian.  The ratio of final mass to initial mass is
  $M_f/M = 0.95162\pm 0.00002$, and the final black hole spin is $S_f/M_f^2=0.68646\pm 0.00004$.

\end{abstract}


\maketitle


\section{Introduction}
\label{sec:Introduction}

Beginning with the groundbreaking binary black hole evolutions of
Pretorius~\cite{Pretorius2005a} and the development of the moving
puncture method~\cite{Campanelli2006a, Baker2006a}, it has recently
become possible to solve Einstein's equations numerically for the
inspiral, merger, and ringdown of two black holes in a binary orbit.
Already these simulations have provided tests of post-Newtonian
approximations~\cite{Buonanno-Cook-Pretorius:2007,Baker2006d,Pan2007,Buonanno2007,Hannam2007,Boyle2007,Gopakumar:2007vh,Hannam2007c,Boyle2008a,Mroue2008,Hinder2008b}, 
have allowed initial exploration of
the orbital dynamics of spinning binaries~\cite{Campanelli2006c,Campanelli2006d,Campanelli2007b,Herrmann2007c,MarronettiEtAl:2008,Berti2007b}, have
determined the recoil velocity of the final black hole when the masses
are unequal~\cite{Campanelli2005,Herrmann2007b,Baker2006c,Gonzalez2007},
and have led to the discovery of dramatically large recoil velocity
from certain spin configurations~\cite{Campanelli2007,Gonzalez2007b,Bruegmann-Gonzalez-Hannam-etal:2007,Herrmann2007,Herrmann2007c,Choi-Kelly-Boggs-etal:2007,Baker2007,Tichy:2007hk,Schnittman2007,Campanelli2007a,Koppitz2007,MillerMatzner2008,Baker2008,Healy2008}.

Waveforms from these numerical simulations are important for
gravitational-wave detectors such as LIGO and LISA.  This is not only
because detected waveforms can be compared with numerical models to
measure astrophysical properties of the sources of gravitational
radiation, but also because the detection probability itself can be
increased via the technique of matched
filtering~\cite{Finn1992}, in which noisy data are convolved
with numerical templates to enhance the signal.

However, binary black hole simulations are time consuming: a single
simulation following approximately 10 orbits, merger, and ringdown
typically requires a few weeks of runtime on approximately 50 or 100
processors of a parallel supercomputer, and typically such a
simulation produces waveforms of only modest accuracy.  This
large computational expense precludes, for example, producing a full
template bank of numerical waveforms covering the entire parameter
space of black hole masses and spins.  Hence there has been much
interest in construction of phenomenological analytical
waveforms~\cite{Ajith-Babak-Chen-etal:2007,Buonanno2007,Damour2007a,DN2008}
that can be computed
quickly and are calibrated by
a small number of numerical simulations.  While the accuracy of
typical simulations is sufficient for creating LIGO detection
templates, it is most likely inadequate for LIGO
parameter estimation and is far from what is required
for LISA data analysis~\cite{Lindblom2008}.

One approach to increasing the accuracy and efficiency of simulations
is to adopt more efficient numerical methods.  In particular, a class
of numerical techniques known as spectral methods holds much promise.
For smooth solutions, the errors produced by spectral methods decrease
exponentially as computational resources are increased, whereas the
errors of finite difference methods, the methods used by the majority
of binary black hole simulations, decrease polynomially.  Indeed,
spectral methods have been used to produce very accurate initial
data for binary black holes and neutron stars~\cite{Bonazzola1996,Bonazzola1999a,grandclement-etal-2001,Gourgoulhon2001,Grandclement2002,Pfeiffer2002a,Pfeiffer2003,Cook2004,AnsorgBruegmann2004,Ansorg:2005,Caudill-etal:2006,Grandclement2006,Lovelace2008,FoucartEtAl:2008},
and they have been used to produce the longest and most accurate
binary black hole inspiral simulation to date~\cite{Scheel2006,Boyle2007}.

However, a key difficulty with time-dependent spectral binary black
hole simulations has been handling the merger of the two holes.  For
example, the spectral simulations described
in~\cite{Scheel2006,Boyle2007,Boyle2008a} are very accurate and
efficient, but they follow only the inspiral of the two black holes,
and fail just before the holes merge.  This is sufficient for some
applications, such as comparing post-Newtonian formulae with numerical
results during the inspiral and finding accurate analytic templates
that match the numerical inspiral
waveforms~\cite{Boyle2007,Boyle2008a}, but for most purposes the
merger is the most crucial part of the process: for instance the
gravitational-wave emission is the strongest during merger, and
details of the merger determine the recoil velocity of the final black
hole.

In this paper we present a spectral binary black hole simulation that
follows sixteen orbits of the binary plus merger and ringdown of the
merged black hole.  In Sec.~\ref{sec:NumericalSpacetime} we
describe the equations, gauge conditions, and numerical methods we use
to solve Einstein's equations; in particular,
Secs.~\ref{sec:MergerEvolution} and~\ref{sec:RingdownEvolution}
describe changes to our gauge conditions that allow simulation of the
merger, and our method for extending the evolution through ringdown.
In Sec.~\ref{sec:Waveform} we discuss extraction of the
gravitational waveform from the simulation, including the process of
extrapolating the waveform to infinity.  Sec.~\ref{sec:Waveform}
also includes an estimate of the uncertainty in the waveform from
several sources.  Finally, in Sec.~\ref{sec:Discussion} we discuss
outstanding difficulties and future improvements.

\section{Solution of Einstein's Equations}
\label{sec:NumericalSpacetime}

\subsection{Initial data}
\label{sec:InitialData}

The initial data describe two nonspinning black holes, each with
Christodoulou mass $M/2$, in quasicircular orbit with low
eccentricity.  The initial data are exactly as described in
Ref.~\cite{Boyle2007}.  Briefly, initial data are constructed within
the conformal thin sandwich formalism~\cite{York1999,Pfeiffer2003b}
using a pseudospectral elliptic solver~\cite{Pfeiffer2003}.  We
employ quasiequilibrium boundary conditions~\cite{Cook2002,Cook2004}
on spherical excision boundaries, choose conformal flatness and
maximal slicing, and use Eq.~(33a) of Ref.~\cite{Caudill-etal:2006} as
the lapse boundary condition.  The spins of the black holes are made
very small ($\sim 10^{-7}$) via an appropriate choice of the tangential
shift at the excision surfaces, as described
in~\cite{Caudill-etal:2006}.  Finally, the initial orbital
eccentricity is tuned to a very small value ($\sim 5\times 10^{-5}$) using the
iterative procedure described in Ref.~\cite{Boyle2007}, which is an
improved version of the procedure of
Ref.~\cite{Pfeiffer-Brown-etal:2007}.

\subsection{Evolution of the inspiral phase}
\label{sec:InspiralEvolution}

The evolution of the first $\sim 15$ binary orbits is identical
to the simulation presented in Ref.~\cite{Boyle2007}. We describe it
here briefly in order to facilitate the presentation of our method
for continuing the evolution through merger and ringdown, which is
described in
Secs.~\ref{sec:MergerEvolution} and~\ref{sec:RingdownEvolution}.

The Einstein evolution equations are solved with the pseudospectral
evolution code described in Ref.~\cite{Scheel2006}.  This code evolves
a first-order representation~\cite{Lindblom2006} of the generalized
harmonic system~\cite{Friedrich1985,Garfinkle2002,Pretorius2005c}.  
We handle the singularities by excising the black hole
interiors from the computational domain. Our outer boundary
conditions~\cite{Lindblom2006,Rinne2006,Rinne2007} are designed to
prevent the influx of unphysical constraint
violations~\cite{Stewart1998,FriedrichNagy1999,Bardeen2002,Szilagyi2002,%
Calabrese2003,Szilagyi2003,Kidder2005}
and undesired incoming gravitational radiation~\cite{Buchman2006,Buchman2007},
while allowing the outgoing gravitational radiation to pass freely
through the boundary.

We employ the dual-frame method described in Ref.~\cite{Scheel2006}:
we solve the equations in an ``inertial frame'' that is asymptotically
Minkowski, but our domain decomposition is fixed in a ``comoving frame''
that rotates with respect to the inertial frame and also shrinks with
respect to the inertial frame as the holes approach each other. The
positions of the holes are fixed in the comoving frame; we account for
the motion of the holes by dynamically adjusting the coordinate
mapping between the two frames.  Note that the comoving frame is
referenced only internally in the code as a means of treating moving
holes with a fixed domain. Therefore all coordinate quantities
(e.g. black hole trajectories, wave-extraction radii) mentioned in
this paper are inertial-frame values unless explicitly stated otherwise.

As described in~\cite{Boyle2007}, the mapping between inertial and
comoving coordinates for the inspiral, expressed in polar coordinates
relative to the center of mass of the system, is
\begin{eqnarray}
\label{eq:CubicScaleMap}
r      &=& \left[a(t) + \left(1-a(t)\right) \frac{r'^2}{R_0'^2} \right] r', \\
\theta &=& \theta',\\
\phi   &=& \phi' + b(t),\label{eq:CubicScaleMapPhi}
\end{eqnarray}
where $a(t)$ and $b(t)$ are functions of time, and $R_0'$ is a
constant usually chosen to be roughly the radius of the outer boundary in
comoving coordinates.  Here primes denote the comoving coordinates.
For the choice $R_0'=\infty$, the mapping is simply a rotation by $b(t)$
plus an overall contraction given by $a(t)$.
The functions $a(t)$ and $b(t)$ are determined by a dynamical control
system as described in Ref.~\cite{Scheel2006}.  This control system
dynamically adjusts $a(t)$ and $b(t)$ so that the centers of the
apparent horizons remain stationary in the comoving frame.  Note that
the outer boundary of the computational domain is at a fixed comoving
radius $R'_{\rm max}$, so the inertial-coordinate radius of the outer
boundary $R_{\rm max}(t)$ is a function of time.

The gauge freedom in the generalized harmonic system is fixed
via a freely specifiable gauge source function $H_a$ that satisfies the
constraint 
\begin{equation}
  \label{e:ghconstr}
  0 = \mathcal{C}_a \equiv \Gamma_{ab}{}^b + H_a,
\end{equation}
where $\Gamma^{a}{}_{bc}$ are the spacetime Christoffel symbols.
To choose this gauge source function, we first define a new
quantity $\tilde{H}_a$ that has the following two properties: (1)
$\tilde{H}_a$ transforms like a tensor, and (2) in inertial coordinates
$\tilde{H}_a = H_a$.  We choose $H_a$ so that the constraint
equation~(\ref{e:ghconstr}) is satisfied initially, and we demand that
$\tilde{H}_{a'}$ is constant in the moving frame, {\it i.e.,} that
$\partial_{t'} \tilde{H}_{a'} = 0$.

\subsection{Extending inspiral runs through merger}
\label{sec:MergerEvolution}

If the inspiral runs described above are allowed to continue without
any modification of the algorithm, then as the binary approaches
merger, the horizons of the black holes become extremely distorted and
the dynamical fields begin to develop sharp (but numerically
convergent) features near each hole.  These features grow rapidly in
time, eventually halting the simulation before merger. This is due to
a gauge effect: The gauge condition used during the inspiral, namely
fixing $H_a$ in time in the comoving frame, was chosen based on the
idea that each black hole is in quasiequilibrium in this frame. Once
the black holes begin to interact strongly, this gauge condition no
longer allows the coordinates to sufficiently react to the changing
geometry, and coordinate singularities develop.

Therefore we must modify our gauge conditions in order to handle
merger.  Because the inspiral gauge works so well before merger,
we choose to remain in that gauge until some time $t=t_g$, and then we
change (smoothly) to a new gauge.  

We have experimented with several gauge conditions~\cite{Lindblom2007}, but so
far the simplest gauge choice that works, and the one used in the
simulations presented here, is based on the gauge treatment of
Pretorius~\cite{Pretorius2005c,Pretorius2005a,Pretorius2006}:
We promote the gauge source function $H_a$ to an independent dynamical
field that satisfies
\begin{equation}
\label{eq:Hevolution}
\nabla^c\nabla_c H_a = Q_a(x,t,\psi_{ab}) + \xi_2 t^b\partial_b H_a,
\end{equation}
where $\nabla^c\nabla_c$ is the curved space {\it scalar\/} wave
operator (i.e. each component of $H_a$ is evolved as a scalar),
$\psi_{ab}$ is the spacetime metric, and $t^a$ is the timelike unit normal
to the hypersurface.
The driving function $Q_a$ is
\begin{eqnarray}
\label{eq:Hevolutiont}
Q_t &=& f(x,t)\xi_1\frac{1-N}{N^\eta},\\
\label{eq:Hevolutioni}
Q_i &=& g(x,t)\xi_3\frac{N_i}{N^2}.
\end{eqnarray}
Here $N$ and $N^i$ are the lapse function and the shift vector,
$\eta$, $\xi_1$, $\xi_2$, and $\xi_3$ are constants, and $f(x,t)$
and $g(x,t)$ are prescribed functions of the spacetime coordinates (we
describe our choices for these objects below). 

Equation~(\ref{eq:Hevolution}) is a damped, driven wave equation with
damping parameter $\xi_2$ and driving function $Q_a$.  The driving
term $Q_t$ in Eq.~(\ref{eq:Hevolutiont}) was introduced by
Pretorius~\cite{Pretorius2005a,Pretorius2005c} to drive the lapse
function towards unity so as to prevent it from becoming small.  The
driving term $Q_i$ is new; it drives the shift vector towards zero
near the horizons.  This causes the horizons to expand in
coordinate space, and has the effect of smoothing out the dynamical
fields near the horizon and preventing gauge singularities from
developing.  A different gauge choice that causes similar coordinate
expansion of the horizons was introduced in Ref.~\cite{Szilagyi2007}.
Care must be taken so that the horizons do not expand
too quickly relative to the excision boundaries; otherwise the
characteristic fields will fail to be purely outgoing (into the
holes) at the excision boundaries, and excision will fail. We find
that with appropriate choices of $\xi_1$, $\xi_3$, $f(x,t)$, and $g(x,t)$
as described below, the horizons expand gradually and not too rapidly.

For the runs presented here we choose $\eta=4$, $\xi_1=0.1$,
$\xi_2=10$, and $\xi_3=0.4$. The functions $f(x,t)$ and $g(x,t)$ in
Eqs.~(\ref{eq:Hevolutiont}) and~(\ref{eq:Hevolutioni}) are chosen
based on two criteria: the first is that the driving terms $Q_a$ are
nonzero only near the black holes where they are needed; if these
terms are nonzero in the wave-extraction zone they lead to complicated
gauge dynamics in this region, making waveform extraction difficult.  The
second criterion is that the driving terms are turned on 
in a gradual manner so that the gauge does not change too rapidly.  We
choose
\begin{eqnarray}
  f(x,t) = g(x,t) &=& (2-e^{-(t-t_g)/\sigma_1}) \nonumber\\
  &\times& (1-e^{-(t-t_g)^2/\sigma_2^2}) 
  e^{-r'^2/\sigma_3^2},\label{eq:GaugeRolloff}
\end{eqnarray}
where $r'$ is the coordinate radius in comoving coordinates, and the
constants are $\sigma_1\sim 17.5M$, $\sigma_2\sim 15M$, and $\sigma_3\sim 40M$.
Here $M$ is the sum of the initial Christodoulou masses of the two
holes.

Equation~(\ref{eq:Hevolution}) is a second-order hyperbolic equation, which
we evolve in first-order form by defining
new fields $\Pi^H_a$ and $\Phi^H_{ia}$, representing (up to the
addition of constraints) the appropriate time and space derivatives of
$H_a$, respectively:
\begin{eqnarray}
  \Pi^H_a &=& -t^b \partial_b H_a, \\ 
  \Phi^H_{ia} &=& \partial_i H_a .
\end{eqnarray}
The representation of wave equations of this type in
first-order form is well understood, see e.g., Refs.~\cite{Holst2004,%
Lindblom2006}; the result for Eq.~(\ref{eq:Hevolution}) is
\begin{eqnarray}
\partial_t H_a &=& -N \Pi^{H}_a + N^k\Phi^{H}_{ka},\label{e:psidot}\\
\partial_t \Pi^{H}_a &=& N^k\partial_k \Pi^H_a - Ng^{ki}\partial_k \Phi^H_{ia}
-\gammaHtwo N^k\partial_kH_a \nonumber\\
&+&\gammaHtwo N^k\Phi^H_{ka}+
N (\Gamma^{kj}{}_j-g^{kj}\partial_j N) \Phi^H_{ka} \nonumber\\
&+& NK \Pi^H_a+ Q_a,
\label{e:pidot}\\
\partial_t \Phi^H_{ia} &=& N^k\partial_k \Phi^H_{ia} - N\partial_i \Pi^H_a 
+\gammaHtwo N\partial_i H_a \nonumber\\
&-& \Pi^H_a\partial_iN+ \Phi^H_{ka}\partial_iN^k
-\gammaHtwo N \Phi^H_{ia},\label{e:Phidot}
\end{eqnarray}
where $g_{ij}$ is the spatial metric and
$K$ is the trace of the extrinsic curvature.
We choose the constraint-damping parameter $\gammaHtwo$ to be 
$\gammaHtwo=4/M$.

These equations are symmetric hyperbolic, and require boundary
conditions on all incoming characteristic fields at all boundaries.
The characteristic fields for Eqs.~(\ref{e:psidot})--(\ref{e:Phidot})
in the direction of a unit spacelike covector $n_i$ are
\begin{eqnarray}
\label{eq:CharFieldsU}
U^{H\pm}_a  &=& \Pi^H_a\pm n^i \Phi^H_{ia}-\gammaHtwo H_a,\\
\label{eq:CharFieldsZ1}
Z^{H1}_a    &=& H_a,\\
\label{eq:CharFieldsZ2}
Z^{H2}_{ia} &=& (\delta_i^k - n_i n^k) \Phi^H_{ka}.
\end{eqnarray}
The (coordinate) characteristic speeds for $U^{H\pm}_a$, $Z^{H1}_a$,
and $Z^{H2}_{ia}$ are $\pm N - n_i N^i$, $0$, and $-n_iN^i$,
respectively.

At the excision boundaries all characteristic fields are outgoing
(i.e. into the holes) or nonpropagating, so no boundary conditions are
necessary and none are imposed.  At the outer boundary, we must impose
boundary conditions on $U^{H-}_a$ and $Z^{H2}_{ia}$.  Define
\begin{eqnarray}
D_t(U^{H\pm}_a)  &\equiv& \partial_t\Pi^H_a\pm n^i \partial_t\Phi^H_{ia}
                       -\gammaHtwo \partial_t H_a,\\
D_t(Z^{H1}_a)    &\equiv& \partial_t H_a,\\
D_t(Z^{H2}_{ia}) &\equiv& (\delta_i^k - n_i n^k) \partial_t\Phi^H_{ka},
\end{eqnarray}
where the time derivatives on the right-hand side are evaluated
using Eqs.~(\ref{e:psidot})--(\ref{e:Phidot}).
Then we impose the following boundary conditions:
\begin{eqnarray}
\label{eq:BcU-}
\partial_t U^{H-}_a   &=& - \gammaHtwo D_t(Z^{H1}_a),\\
\label{eq:BcZ2}
\partial_t Z^{H2}_{ia} &=& D_t(Z^{H2}_{ia})+2n_kN^kn^j\partial_{[i} \Phi^H_{j]a}.
\end{eqnarray}
Equation~(\ref{eq:BcU-}) is the outgoing-wave boundary condition described
in detail in Ref.~\cite{Holst2004}. Equation~(\ref{eq:BcZ2}) ensures that
violations of the artificial constraint
$C_{ia}\equiv\Phi^H_{ia}-\partial_i H_a=0$ do not enter the domain
through the boundary; it is the direct analogue of the
constraint-preserving boundary condition we apply to the analogous
variable in the generalized harmonic formulation of Einstein's
equations, Eq.~(65) of Ref.~\cite{Lindblom2006}.

Note that Eqs.~(\ref{e:psidot})--(\ref{e:Phidot}) involve only first
derivatives of the spacetime metric, and similarly, the generalized
harmonic Einstein equations involve only first derivatives of $H_a$.
Therefore, adding Eqs.~(\ref{e:psidot})--(\ref{e:Phidot}) to the
system does not change the hyperbolicity or characteristic fields of
the generalized harmonic Einstein equations, so we can impose the
same boundary conditions on the generalized harmonic variables as we
do during the inspiral, as described in Refs.~\cite{Scheel2006,Rinne2006}.

Equations~(\ref{e:psidot})--(\ref{e:Phidot}) require as initial data the
values of $H_a$ and $\Pi^{H}_a$ at $t=t_g$.  These quantities can be
computed from the gauge choice used during the inspiral for $t\leq
t_g$, so we choose them to be continuous at $t=t_g$.

Note that Eqs.~(\ref{e:psidot})--(\ref{e:Phidot}) and the boundary
conditions~(\ref{eq:BcU-}) and~(\ref{eq:BcZ2}) are written in the
inertial coordinate system. The equations are actually solved in the
comoving coordinate system using the dual-frame method described in
Ref.~\cite{Scheel2006}.

With the modifications to the gauge conditions described here, the
evolution of the binary can be tracked up until (and shortly after)
the formation of a common horizon that encompasses both black holes.
Because of the more rapid dynamics and the distortions of the horizons
during the merger, we typically increase the numerical resolution
slightly when we make these changes to the gauge conditions (this is
the difference between the first and second entry in the $N_{\rm pts}$
column in Table~\ref{tab:Evolutions}).  After the common horizon
forms, the problem reduces to evolving a single highly distorted
dynamical black hole, rather than two separate black holes.  We change
the algorithm to take advantage of this, as described in the next
section.

\subsection{Evolution from merger through ringdown}
\label{sec:RingdownEvolution}

We make three main changes to our evolution algorithm once we detect a common
apparent horizon.  First, because there is now only one black hole and
not two, we interpolate all variables onto a new computational domain
that contains only a single excised region.  Second, we choose a new
comoving coordinate system (and a corresponding mapping to inertial
coordinates) so that the new excision boundary tracks the shape of the
(distorted, rotating, pulsating) apparent horizon in the inertial
frame, and so that the outer boundary behaves smoothly in time.
Third, we modify the gauge conditions so that the shift vector is no
longer driven towards zero, allowing the solution to eventually relax
to a time-independent state. We now discuss these three changes in
detail.

Our new computational domain contains only a single excised region,
and is much simpler than the one used until merger. It consists only
of nested spherical-shell subdomains that extend from a new excision
boundary $R''_{\rm min}$, chosen to be slightly inside the common
apparent horizon, to an outer boundary $R''_{\rm max}$ that coincides
with the outer boundary of the old domain. 

To understand how we choose our new comoving frame, first recall that
in the dual-frame technique~\cite{Scheel2006}, the comoving frame is
the one in which the computational domain is fixed, the inertial frame
is the one in which the coordinates are Minkowski-like at infinity,
and the two frames are related by a mapping that is chosen so that the
computational domain tracks the motion of the black holes.  Let $x^a$
represent the inertial coordinates (which are the same before and
after merger), let $x'^a$ represent the old comoving coordinates, and
let $x''^a$ represent the new comoving coordinates.  The mapping
between $x'^a$ and $x^a$ is given by
Eqs.~(\ref{eq:CubicScaleMap})--(\ref{eq:CubicScaleMapPhi}).  The
mapping between $x''^a$ and $x^a$ is chosen to be
\begin{eqnarray}
\label{eq:PostMergerMap}
r         &=&  \tilde{r}\biggr[1+\sin^2(\pi\tilde{r}/2 R''_{\rm max})  
                    \nonumber \\
                    &&\left. \times
                    \left(A(t)\frac{R'_{max}}{R_{\rm max}''} 
                       + (1-A(t))\frac{R_{max}'^3}{R_{\rm max}''R_0'^2} 
                       -1\right) \right], \\
\label{eq:PostMergerMapDistort}
\tilde{r} &=& r'' - q(r'')\sum_{\ell=0}^{\ell_{\rm max}} 
                       \sum_{m=-\ell}^{\ell} \lambda_{\ell m}(t)
                       Y_{\ell m}(\theta'',\phi''),\\
\theta &=& \theta'',\\
\phi   &=& \phi'' + B(t),\label{eq:PostMergerMapPhi}
\end{eqnarray}
where $R'_{max}$ is the outer boundary of the premerger computational
domain in the old comoving coordinates, and $q(r'')$, $A(t)$, $B(t)$,
and $\lambda_{\ell m}(t)$ are functions we will now discuss.

First we describe the angular map: The function $B(t)$ is chosen so that the new
comoving frame initially rotates with respect to the inertial frame,
but this rotation slows to a halt after a short time. In particular,
\begin{equation}
B(t) = B_0 + (B_1+B_2(t-t_m))e^{-(t-t_m)/\tau_B},
\end{equation}
where the constants $B_0$, $B_1$, and $B_2$ are chosen so that
$B(t)$ matches smoothly onto $b(t)$ from Eq.~(\ref{eq:CubicScaleMapPhi}):
$B(t_m)=b(t_m)$, $\dot{B}(t_m)=\dot{b}(t_m)$, and $\ddot{B}(t_m)=\ddot{b}(t_m)$.
Here $t_m$ is the time at which we transition to the new domain decomposition.
The constant $\tau_B$ is chosen to be on the order of $20M$.

The radial map is a composition of two individual maps:
Eqs.~(\ref{eq:PostMergerMap}) and~(\ref{eq:PostMergerMapDistort}).
The purpose of Eq.~(\ref{eq:PostMergerMap}) is to match the outer
boundary of the new domain smoothly onto that of the old domain, while
far from the outer boundary Eq.~(\ref{eq:PostMergerMap}) approaches
the identity.  We have found that without the use of
Eq.~(\ref{eq:PostMergerMap}), the (inertial-coordinate) location of
the boundary changes nonsmoothly at $t=t_m$, thereby generating a
spurious ingoing gauge pulse that spoils waveform extraction.
The function $A(t)$ is 
\begin{equation}
A(t) = A_0 + (A_1+A_2(t-t_m))e^{-(t-t_m)/\tau_A},
\end{equation}
where the constants $A_0$, $A_1$, and $A_2$ are chosen so that
$A(t)$ matches smoothly onto $a(t)$ from Eq.~(\ref{eq:CubicScaleMap}):
$A(t_m)=a(t_m)$, $\dot{A}(t_m)=\dot{a}(t_m)$, and $\ddot{A}(t_m)=\ddot{a}(t_m)$.
The constant $\tau_A$ is chosen to be on the order of $5M$.

The other piece of the radial map, Eq.~(\ref{eq:PostMergerMapDistort}),
is chosen so that the apparent horizon is nearly
spherical in the new comoving
coordinates $x''^a$. The function $q(r'')$ is
\begin{equation}
q(r'') = e^{-(r''-R''_{\rm AH})^3/\sigma_q^3},
\end{equation}
where $R''_{\rm AH}$ is the radius of the apparent horizon in comoving
coordinates, and $\sigma_q$ is a constant of order $20M$.  This
function $q(r'')$ ensures that the piece of the radial map represented
by Eq.~(\ref{eq:PostMergerMapDistort}) acts only in the vicinity of
the merged hole and not in the exterior wave-extraction region. 

We now discuss the choice of the functions $\lambda_{\ell m}(t)$
that appear in Eq.~(\ref{eq:PostMergerMapDistort}). Given
the known location of the apparent horizon in inertial
coordinates, the $\lambda_{\ell m}(t)$ determine the shape of the
apparent horizon in comoving coordinates. At $t=t_m$, we
choose these
quantities so that the apparent horizon is spherical (up to spherical
harmonic component $\ell=\ell_{\rm max}$) in
comoving coordinates: that is, if the comoving-coordinate
radius of the apparent horizon as
a function of angles is written as
\begin{equation}
\label{eq:AhRadius}
r''_{\rm AH}(\theta'',\phi'') \equiv \sum_{\ell=0}^{\ell_{\rm max}} 
                             \sum_{m=-\ell}^{\ell} Q_{\ell m}(t)
                             Y_{\ell m}(\theta'',\phi''),
\end{equation}
then for $1\le \ell\le \ell_{\rm max}$ we choose $\lambda_{\ell m}(t_m)$ so
that $Q_{\ell m}(t_m)=0$.  In addition, we choose $\lambda_{00}(t_m)=0$;
this determines $R''_{\rm AH}$. For $t>t_m$, $\lambda_{\ell m}(t)$ are
determined by a dynamical feedback control system identical to the one
described in Ref.~\cite{Scheel2006}, which adjusts these functions so
that the apparent horizon is driven to a sphere (up to spherical
harmonic component $\ell=\ell_{\rm max}$) in comoving
coordinates.  This dynamical feedback control allows us to freely
choose the first and second time derivatives of $\lambda_{\ell m}$ at
$t=t_m$.  Simply choosing these to be zero causes the control system
to oscillate wildly before settling down, and unless the time step is
very small, these oscillations are large enough that the excision
boundary crosses the horizon and our excision algorithm fails.  So
instead, we obtain the time derivatives of $\lambda_{\ell m}$ by
finding the apparent horizon at several times surrounding $t=t_m$,
computing $\lambda_{\ell m}$ at these times, and finite-differencing
in time.  For the equal-mass zero-spin merger presented here,
in Eq.~(\ref{eq:PostMergerMapDistort}) it suffices to sum only over
even $\ell$ and $m$ and to choose $\ell_{\rm max}=6$.

The last change we make before continuing the simulation past merger
is to modify the functions
$f(x,t)$ and $g(x,t)$, which before merger were given by
Eq.~(\ref{eq:GaugeRolloff}), to
\begin{eqnarray}
  f(x,t) &=& (2-e^{-(t-t_g)/\sigma_1}) \nonumber\\
  &\times& (1-e^{-(t-t_g)^2/\sigma_2^2}) 
  e^{-r''^2/\sigma_3^2},\label{eq:GaugeRolloff2} \\
  g(x,t) &=& f(x,t)e^{-(t-t_m)^2/\sigma_4^2},
\end{eqnarray}
where $\sigma_4=7M$.  The modification of $g(x,t)$ turns off the term
in the gauge evolution equations that drives the shift to zero near
the holes.  Before merger, it is advantageous to have the shift driven
to zero so that the horizons expand in coordinate space and so that
growing gauge modes remain inside the common horizon. 
After merger, however, it is no longer desirable for the
horizon to expand, since this would prevent the solution from eventually
settling down to a time-independent state in which the horizon is
stationary with respect to the coordinates.

To summarize, the steps involved in the transition from evolving a
binary black hole spacetime to evolving a merged single black hole
spacetime are as follows: (1) Find the common apparent horizon in the
inertial frame at several times near $t=t_m$. (2) Solve for the
$\lambda_{\ell m}(t_m)$ that make the horizon spherical in the
comoving frame, and simultaneously solve for $R''_{\rm AH}$.  (3)
Choose the inner boundary of the new computational domain $R''_{\rm
  min}$ to be slightly less than $R''_{\rm AH}$, and choose the outer
boundary $R''_{\rm max}$ [for sufficiently small $a(t_m)$ it is necessary
to choose $R''_{\rm max} < R'_{\rm max}$ so that the
mapping~(\ref{eq:PostMergerMap}) is invertible]. At this point the
computational domain and the
mapping~(\ref{eq:PostMergerMap})--~(\ref{eq:PostMergerMapPhi}) have
been determined. (4) Interpolate all dynamical variables from the old
computational domain onto the new one. This interpolation is done via
the spectral expansion in the old domain, so it introduces no
additional error.  (5) Modify the gauge source evolution equations so
that the shift is no longer driven to zero. (6) Continue the evolution
on the new computational domain.  All of these steps can be automated.

\subsection{Properties of the numerical solution}
\label{sec:PropertiesOfSolution}

\begin{table}
\begin{tabular}{|c|cccccc|}\hline
Run    & $R'_{\rm max}$ & $R''_{\rm max}$ & $R'_0$ 
& $N_{\rm pts}$  & CPU-h & CPU-h/T \\ \hline
30c1/N4& 462&462&698     & $(57^3,59^3,57^3)$& 8,800  & 2.0 \\
30c1/N5& 462&462&698     & $(62^3,66^3,63^3)$& 15,000 & 3.4 \\
30c1/N6& 462&462&698     & $(67^3,73^3,70^3)$& 23,000 & 5.3 \\
30c2/N6& 722&96 &$\infty$& $(71^3,76^3,63^3)$& 25,000 & 5.7 \\
\hline
\end{tabular}
\caption{Outer boundary parameters,
  collocation points, and CPU usage for several zero-spin binary black hole
  evolutions.  The first column identifies the inspiral run in the
  nomenclature of Ref.~\cite{Boyle2007}.  $N_{\rm pts}$ is the
  approximate number of collocation points used to cover the entire
  computational domain.  The three values for $N_{\rm pts}$ are those
  for the inspiral, merger, and ringdown portions of the
  simulation, which are described in Sections~\ref{sec:InspiralEvolution},
  \ref{sec:MergerEvolution}, and~\ref{sec:RingdownEvolution}, respectively.
  The outer boundary parameters $R'_{\rm max}$, $R''_{\rm max}$ and $R'_0$,
  as well as run times $T$,
  are in units of the initial Christodoulou mass $M$ of the system,
  which provides a natural time and length scale.  
  \label{tab:Evolutions}
}
\end{table}

\begin{figure}
\includegraphics[width=\linewidth]{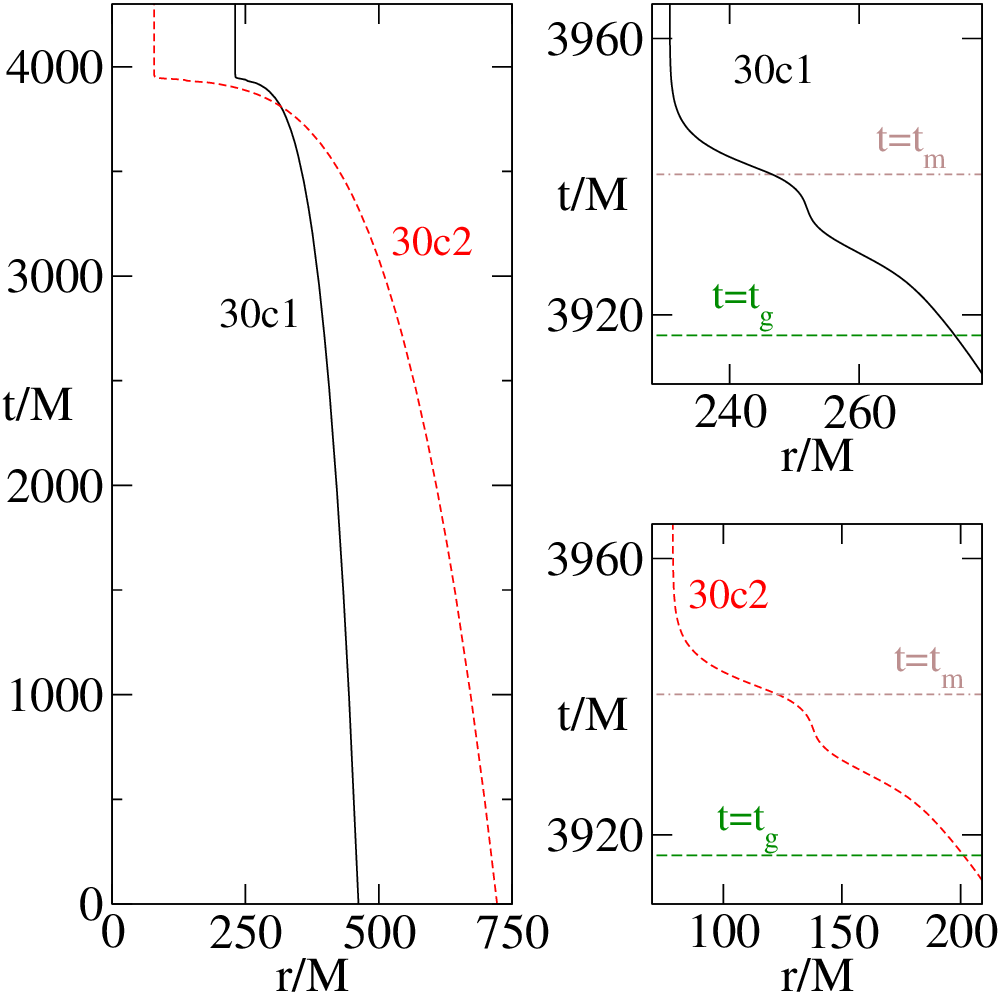}
\caption{\label{fig:SpacetimeDiagram}Spacetime diagram showing the
  spacetime volume simulated by the numerical evolutions listed in
  Table~\ref{tab:Evolutions}.  Each curve represents the worldline of
  the outer boundary for a particular simulation. The magnified views
  on the right show that the outer boundary moves smoothly near
  merger. The transition times $t_g=3917M$ and $t_m=3940M$ are indicated
  on the right panels.} 
\end{figure}

In Table~\ref{tab:Evolutions} we list outer boundary parameters, resolutions,
and run times of several runs we
have done using the algorithm described above.  Three of these runs
are identical except for numerical resolution, and the fourth is
performed on a different domain with a different outer boundary
location. As discussed above, the outer boundary of our simulation
varies in time because of the dual-frame approach we use to follow the
black holes. Figure~\ref{fig:SpacetimeDiagram} is a spacetime diagram
illustrating the region of spacetime being evolved in our simulation.

\begin{figure}
\includegraphics[width=\linewidth]{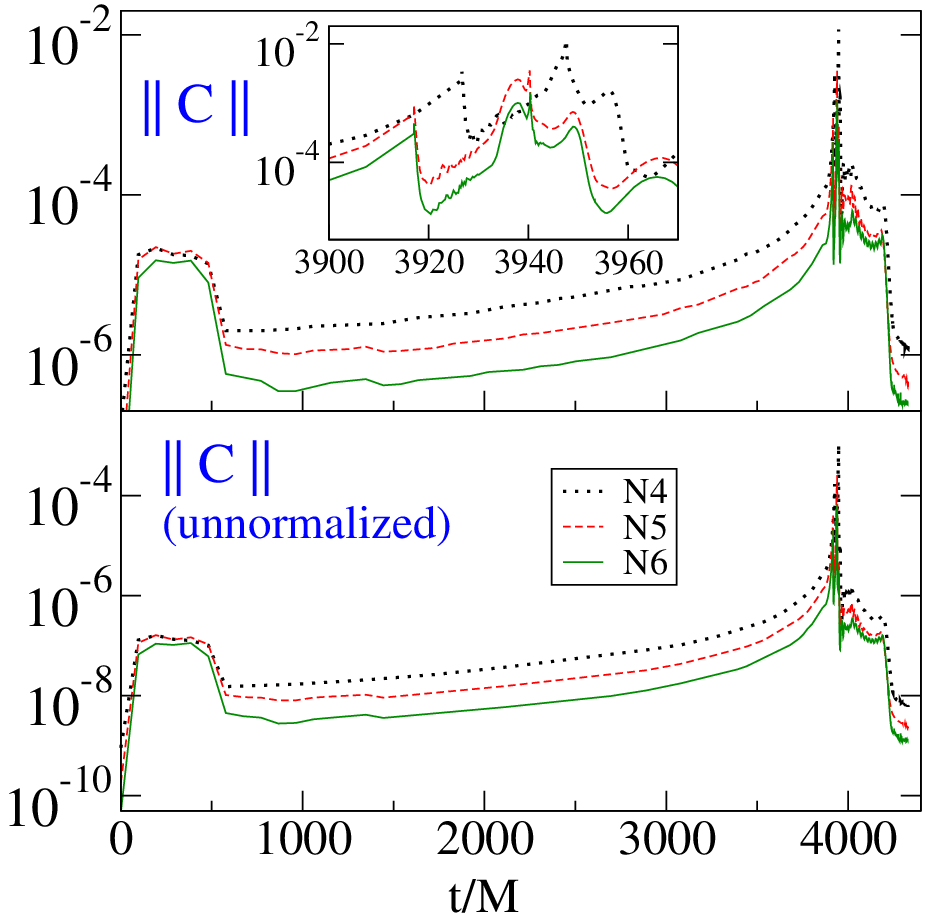}
\caption{\label{fig:Constraints} Constraint violations of run 30c1.
  The top panel shows the
  $L^2$ norm of all constraints, normalized by the $L^2$ norm of
  the spatial gradients of all dynamical fields. 
  The bottom panel shows the same data, but without the normalization factor.
  The $L^2$ norms are taken over the portion of the computational volume
  that lies outside apparent horizons.
}
\end{figure}

We do not explicitly enforce either the Einstein
constraints or the secondary constraints that arise from writing the
system in first-order form.  Therefore, examining how well these
constraints are satisfied provides a useful consistency check.  
Figure~\ref{fig:Constraints} shows the constraint
violations for run 30c1.  The top panel shows the $L^2$ norm of all
the constraint fields of our first-order generalized harmonic system,
normalized by the $L^2$ norm of the spatial gradients of the dynamical
fields (see Eq.~(71) of Ref.~\cite{Lindblom2006}).  The bottom panel
shows the same quantity, but without the normalization factor [{\em
  i.e.,} just the numerator of Eq.~(71) of Ref.~\cite{Lindblom2006}].
The $L^2$ norms are taken over the portion of the computational volume
that lies outside apparent horizons.  
At early times, $t<500M$,
the constraints converge rather slowly with resolution because the
junk radiation contains high frequencies.  Convergence is more rapid
during the smooth inspiral phase, after the junk radiation has exited
through the outer boundary. 

The constraints increase as the holes approach each other and the solution
becomes increasingly distorted.  At $t=3917M$ ($t=3927M$ for
resolution N4), the gauge conditions are changed
(cf. Sec.~\ref{sec:MergerEvolution}) and the resolution is
increased slightly (compare the first and second entry in the $N_{\rm
  pts}$ column in Table~\ref{tab:Evolutions}). Because of the change
of resolution, the constraints drop rapidly by almost two orders of
magnitude, but then they begin to grow again.  The transition to a
single-hole evolution (cf. Sec.~\ref{sec:RingdownEvolution}) occurs
at $t=3940M$ ($t=3948M$ for resolution N4). At this time the
constraint norm drops by about an order of magnitude because the
region in which the largest constraint violations occur---the interior of
the common horizon---is newly excised.  

After the binary proceeds through inspiral, merger, and ringdown, it
settles down to a final stationary black hole.  In our simulation this final
state is not expressed in any standard coordinate system used to describe
Kerr spacetime, but nevertheless the final
mass and spin of the hole can be determined.  The area $A$ of the
apparent horizon provides the irreducible mass of the final black hole,
\begin{equation}
M_{\rm irr} = \sqrt{A/16\pi},
\end{equation}
which we find to be $M_{\rm irr}/M = 0.88433\pm 0.00001$,
where $M$ is the sum of the initial irreducible masses of the black holes.
The uncertainty in $M_{\rm irr}/M$ is determined
from the difference between runs 30c1/N6, 30c1/N5, and 30c2/N6, so
it includes only uncertainties due to numerical resolution and outer
boundary location.  We have verified that the uncertainty due to the
finite resolution of our apparent horizon finder is negligible.

\begin{table}
\begin{tabular}{|lrcl|}\hline
Initial orbital eccentricity: & $e$         &$\sim$& $5\times 10^{-5}$ \\
Initial spin of each hole:    & $S_i/M^2$   &$\lesssim$& $10^{-7}$      \\
Time of evolution:            & $T/M$       &$=$&$4330$                \\
Final Christodoulou mass:     & $M_f/M$     &$=$& $0.95162\pm 0.00002$ \\
Final spin:                   & $S_f/M_f^2$ &$=$& $0.68646\pm 0.00004$ \\
\hline
\end{tabular}
\caption{Physical parameters describing the equal-mass nonspinning binary
  black hole evolutions presented here.  The dimensionful quantity $M$
  is the initial sum of the Christodoulou masses of the black holes.
  Uncertainty estimates include numerical uncertainties
  and the effects of varying the outer boundary location.
  \label{tab:EqualMassBinaryParameters}
}
\end{table}

The final spin $S_f$ of the black hole can be computed by integrating a
quasilocal angular momentum density over the final apparent
horizon~\cite{Cook2007,OwenThesis}. Our implementation of this method
is described in detail in Appendix A of~\cite{Lovelace2008}.
Furthermore, an alternative method of computing the final spin, which is
based on evaluating the extremal values of the 2-dimensional scalar
curvature on the apparent horizon and comparing these values to those
obtained analytically for a Kerr black hole, is also described
in~\cite{Lovelace2008}.  Using these measures, we determine the
dimensionless spin of the final black hole to be $S_f/M_f^2 =
0.68646\pm 0.00004$, where the uncertainty is dominated by the
difference between runs 30c1/N6 and 30c1/N5 rather than by the
differences between different methods of measuring the spin.  Here $M_f$
is the Christodoulou mass of the final black hole,
\begin{equation}\label{eq:Christoudoulou-mass}
  M_f^2 =M_{\rm irr}^2+\frac{S_f^2}{4 M_{\rm irr}^2}.
\end{equation}
We find that the ratio of the final to initial black hole
mass is $M_f/M = 0.95162\pm 0.00002$.  The mass and spin of the final
hole are consistent with those found by other groups~\cite{Campanelli2006a,Campanelli2006b,Baker2006b,Bruegmann2006,Buonanno-Cook-Pretorius:2007}.
Physical parameters describing the evolutions are summarized in Table~\ref{tab:EqualMassBinaryParameters}.

\section{Computation of the waveform}
\label{sec:Waveform}

The numerical solution of Einstein's equations obtained using the
methods described above yields the spacetime metric and its first
derivatives at all points in the computational domain.  In this
section we describe how this solution is used to compute the key
quantity relevant for gravitational-wave observations: the
gravitational waveform as seen by an observer infinitely far from the
source.

\subsection{Waveform extraction}
\label{sec:waveform-extraction}

Gravitational waves are extracted from the simulation on a sphere of
coordinate radius $r$
using the Newman-Penrose scalar $\Psi_4$, following the same procedure as in
Refs.~\cite{Pfeiffer-Brown-etal:2007,Boyle2008}.  To summarize, we
compute
\begin{equation}
\Psi_4 = - C_{\alpha\mu\beta\nu} \ell^\mu \ell^\nu \bar{m}^\alpha\bar{m}^\beta,
\label{eq:Psi4Definition}
\end{equation}
where 
\begin{subequations}
\begin{eqnarray}
  \ell^\mu &=& \frac{1}{\sqrt{2}}(t^\mu - r^\mu),\\
     m^\mu &=& \frac{1}{\sqrt{2} r}
            \left(\frac{\partial}{\partial \theta} 
            +    i\frac{1}{\sin\theta}\frac{\partial}{\partial \phi}\right)^\mu.
  \label{eq:FlatspaceMTetrad}
\end{eqnarray}
\end{subequations}
Here $(r,\theta,\phi)$ denote the standard spherical coordinates in the
inertial frame, $t^\mu$ is the timelike unit normal to the spatial hypersurface,
and $r^\mu$ is the outward-pointing unit normal to the extraction sphere.
We then expand $\Psi_4$
in terms of spin-weighted spherical harmonics of weight $-2$:
\begin{equation}
\Psi_4(t,r,\theta,\phi) 
= \sum_{l m} \Psi_4^{l m}(t,r)\, {}_{-2}Y_{l m}(\theta,\phi),
\label{eq:Psi4Ylm}
\end{equation}
where the $\Psi_4^{l m}$ are expansion coefficients defined by this equation.

Note that our choice of
$m^\mu$ is not exactly null nor exactly of unit magnitude at finite
$r$, as is required by the standard definition.  The resulting
$\Psi_4^{l m}$ computed at finite $r$ will therefore 
disagree with the waveforms observed
at infinity.  Our definition does, however, agree with the standard
definition of $\Psi_4^{l m}$ as $r\to\infty$.  Because we extrapolate
the extracted waves to find the asymptotic radiation field (see
Sec.~\ref{sec:waveform-extrapolation}), these tetrad effects should
not play a role: Relative errors in $\Psi_4^{lm}$ introduced by using
the simple coordinate tetrad fall off like powers of $M/r$, and thus
should vanish after extrapolating to obtain the asymptotic behavior.
More careful treatment of the extraction method---such as those
discussed in Refs.~\cite{Nerozzi2006,Pazos2006,Lehner2007}---may
improve the quality of extrapolation and would be interesting to
explore in the future.

In this paper, we focus on the dominant $(l,m)=(2,2)$ mode.  Following
common practice (see e.g.~\cite{Baker2006b,Bruegmann2006}), we split
the extracted waveform into real phase $\phi$ and real amplitude $A$,
defined by
\begin{equation}\label{eq:A-phi-definition}
\Psi^{22}_4(r,t) = A(r,t)e^{-i\phi(r,t)}.
\end{equation}
The gravitational-wave frequency is given by
\begin{equation}\label{eq:omega-definition}
\omega=\frac{d\phi}{dt}
\end{equation}
The minus sign in Eq.~(\ref{eq:A-phi-definition}) is chosen so that
the phase increases in time and $\omega$ is positive.

\begin{figure}
\includegraphics[width=\linewidth]{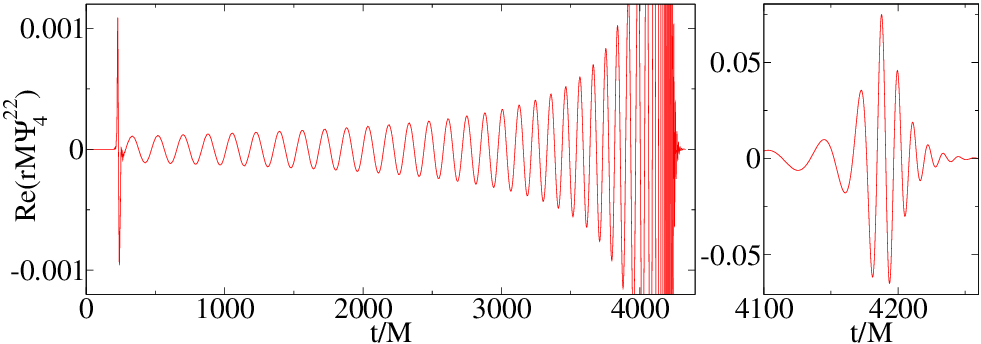}
\caption{\label{fig:Waveform}Gravitational waveform extracted at finite
  radius $r=225M$, for the case 30c1/N6 in Table~\ref{tab:Evolutions}.
  The left panel zooms in on the inspiral waveform,
  and the right panel zooms in on the merger and ringdown.}
\end{figure}

The $(l,m)=(2,2)$ waveform, extracted at a single radius for run
30c1/N6, is shown in Fig.~\ref{fig:Waveform}.  
The short pulse at
$t\sim 200M$ is caused by imperfect initial data that are not
precisely in equilibrium; this pulse is usually referred to as ``junk
radiation''.

\subsection{Convergence of extracted waveforms}
\label{sec:waveform-convergence}

In this section we examine the convergence of the gravitational
waveforms extracted at fixed radius, without extrapolation to
infinity.  This allows us to study the behavior of our code without
the complications of extrapolation.  The extrapolation process and the
resulting extrapolated waveforms are discussed in
Sec.~\ref{sec:waveform-extrapolation}.

\begin{figure}
\includegraphics[width=\linewidth]{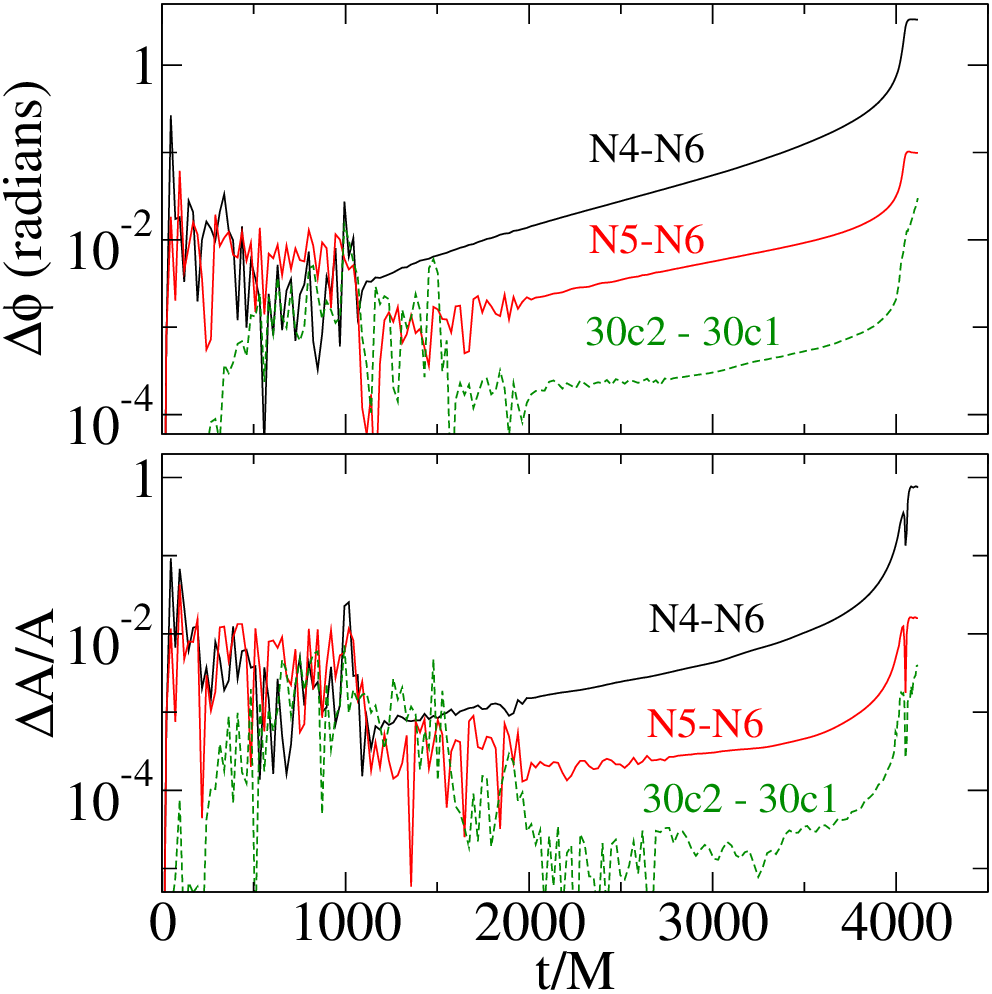}
\caption{\label{fig:WaveformConvergence0} Convergence of waveforms with
  numerical resolution and outer boundary location.  
  Shown are phase and amplitude differences
  between numerical waveforms $\Psi_4^{22}$
  computed using different numerical
  resolutions.  Shown also is the difference between our
  highest-resolution waveforms using two different outer boundary
  locations. All waveforms are extracted at $r=60M$, and no
  time shifting or phase shifting is done to align waveforms.}
\end{figure}

Figure~\ref{fig:WaveformConvergence0} shows the convergence of the
gravitational-wave phase $\phi$ and amplitude $A$ with numerical
resolution.  For this plot, the waveform was extracted at a fixed
inertial-coordinate radius of $r=60M$.
This fairly small extraction radius was chosen to allow a comparison of
the simulations 30c1 and 30c2. Each solid line in the top
panel shows the absolute difference between $\phi$ computed at some
particular resolution and $\phi$ computed from our highest-resolution
run, labeled 30c1/N6 in Table~\ref{tab:Evolutions}. The solid curves
in the bottom panel similarly show the {\it relative\/} amplitude
differences.  When subtracting results at different resolutions, no
time or phase adjustment has been performed.  The noise at early times
is due to ``junk radiation'' generated near $t=0$. 
While most of this radiation leaves through the outer boundary after one
crossing time, some remains visible for a few crossing times\footnote{The
junk radiation at early times is discussed in more detail
in Ref.~\cite{Boyle2007} (specifically, just before Eq. (9) and in
the third paragraph of Sec. II E), which
presents the exact same waveform as shown here but without merger
and ringdown.}. The
plots show that the phase difference accumulated over $16$ orbits plus
merger and ringdown is less than $0.1$ radians for our medium
resolution, and the relative amplitude differences are less than $0.015$;
these numbers can be taken as an estimate of the numerical truncation
error of our {\it medium\/} resolution run. 

Also shown as a dotted curve in each panel of
Fig.~\ref{fig:WaveformConvergence0} is the difference between our
highest-resolution run, 30c1/N6, and a similar run but with a
different outer boundary location, 30c2/N6.  The 30c2 run initially
has a more distant outer boundary than 30c1, but during the inspiral
the outer boundary moves rapidly inward, as seen in
Fig.~\ref{fig:SpacetimeDiagram}, so that extraction of the full
waveform is possible only for extraction radii $r\lesssim 75M$.
Comparing runs 30c1 and 30c2 provides an estimate of the uncertainty
in the waveform due to outer boundary effects such as imperfect
boundary conditions that might reflect outgoing waves. From
Fig.~\ref{fig:WaveformConvergence0} we estimate this uncertainty to be
$0.03$ radians in phase and half a percent in amplitude
(when no time shift is applied).

\begin{figure}
  \includegraphics[width=\linewidth]{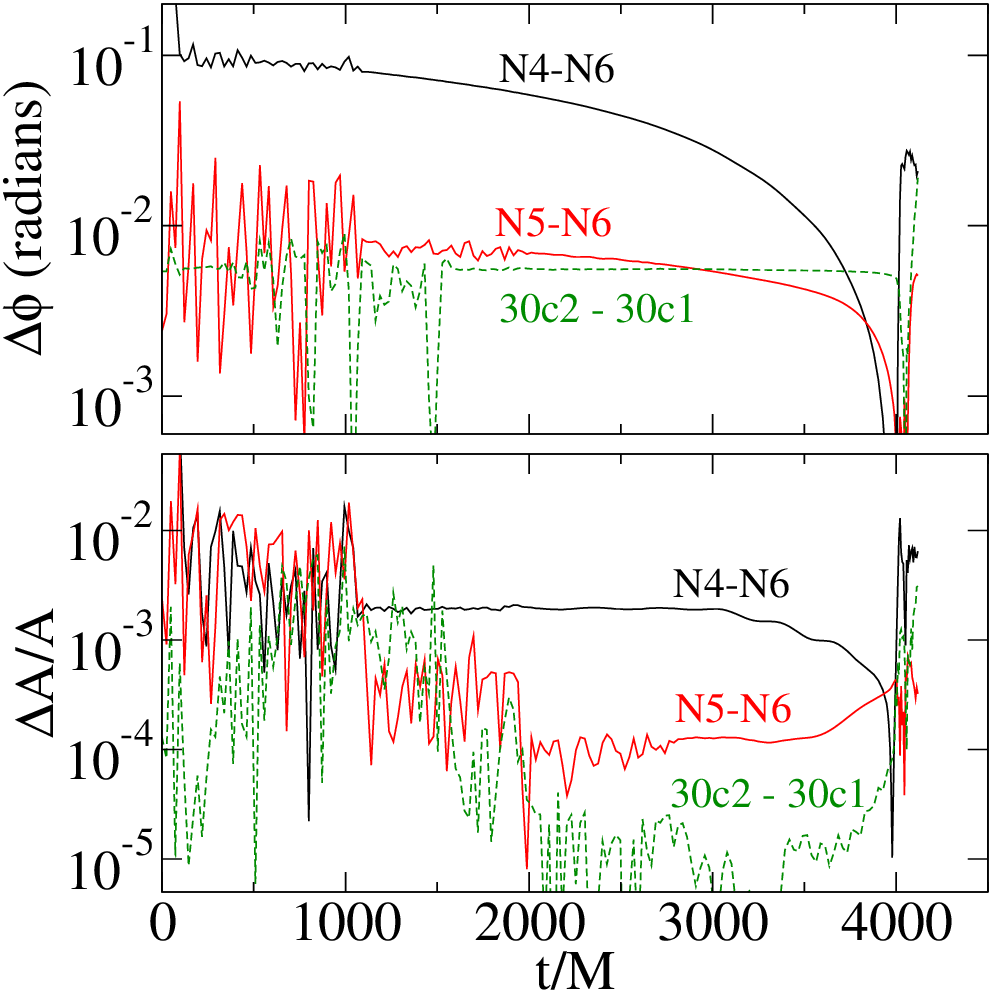}
  \caption{\label{fig:WaveformConvergenceA} Convergence of waveforms
    with numerical resolution and outer boundary location.  Same as
    Fig.~\ref{fig:WaveformConvergence0} except waveforms are
    time-shifted and phase-shifted so that the maximum amplitude
    occurs at the same time and phase.}
\end{figure}

Figure~\ref{fig:WaveformConvergenceA} is the same as
Fig.~\ref{fig:WaveformConvergence0} except each waveform is
time shifted and phase shifted so that the maximum amplitude of the
wave occurs at the same time and phase.  This type of comparison is
relevant for analysis of data from gravitational-wave detectors: when
comparing experimental data with numerical detection templates,
the template will be shifted in both time and phase to best match the data.  
For this type of comparison, Fig.~\ref{fig:WaveformConvergenceA} shows
that the numerical truncation error of our medium
resolution run is less than $0.01$ radians in phase and
$0.1$ percent in amplitude
for $t>1000M$. At earlier times, the errors are somewhat larger and are
dominated by residual junk radiation.  Our uncertainty due to
outer boundary effects is similar to that in
Fig.~\ref{fig:WaveformConvergence0}: about $0.02$ radians in
phase and half a percent in amplitude. Boundary effects are
most prominent during the ringdown.

\subsection{Extrapolation of waveforms to infinity}
\label{sec:waveform-extrapolation}

Our numerical simulations cover only a finite spacetime volume, as
shown in Fig.~\ref{fig:SpacetimeDiagram}, so it is necessary to
extract our numerical waveforms at a finite distance from the source.
However, gravitational-wave detectors measure waveforms as seen by an
observer infinitely far from the source.  Accordingly, after
extracting waveforms at multiple finite radii, we extrapolate these
waveforms to infinite radius using a procedure similar to that
described in~\cite{Boyle2007}.  This extrapolation procedure is
intended to remove not only near-field effects that are absent at
infinity, but also gauge effects that can be caused by
the time-dependence of the lapse function or the nonoptimal choice of
tetrad for computing $\Psi_4$.

The extraction procedure described in
Sec.~\ref{sec:waveform-extraction} yields a set of waveforms
$\Psi_4^{22}(t,r)$, with each waveform extracted at a different
radius.  To extrapolate to infinite radius we must compare waveforms
at different radii, but these waveforms must be offset in time by the
light-travel time between adjacent radii.  To account for this time
shift, for each extraction radius we compute $\Psi_4^{22}(u,r)$, where
$u$ is the retarded time at that radius. 
Assuming for simplicity
that the background spacetime is nearly Schwarzschild,
we compute the retarded time $u$ using
\begin{equation}
\label{eq:RetardedTime}
  u \equiv t_s - r^\ast,
\end{equation}
where $t_s$ is some approximation
of Schwarzschild time, and the tortoise-coordinate radius~\cite{Fiske2005} is
\begin{equation}
\label{eq:Rstar}
  r^\ast = \ra + 
  2\Eadm\ln\left( \frac{\ra}{2\Eadm} - 1 \right).
\end{equation}
Here $\Eadm$ is the ADM mass of the initial data, and
$\ra=\sqrt{A/4\pi}$, where $A$ is the measured (time-dependent) area
of the extraction sphere.  If we were to choose $t_s$ to be
  simply the coordinate time $t$, then the retarded time coordinate
  $u$ would fail to be null, largely because the lapse function in our
  simulation is time-dependent and differs from the Schwarzschild
  value.  We attempt to account for this by assuming that our
  background spacetime coordinates are Schwarzschild, but with
  $g_{tt}$ replaced by $-N_{\rm avg}^2$, where $N_{\rm avg}$ is the
  (time-dependent) average value of the lapse function measured
    on the extraction sphere.  Under these assumptions, it can be
  shown that the one-form
\begin{equation}
\label{eq:NullOneForm}
  \frac{N_{\rm avg}}{\sqrt{1-2\Eadm/\ra}}dt-dr^\ast
\end{equation}
is null, so we equate this one-form with $du$ and thus define
\begin{equation}
\label{eq:SchwarzschildTime}
  t_s = \int_0^t 
  \frac{N_{\rm avg}}{\sqrt{1-2\Eadm/\ra}} dt.
\end{equation}
We show below (cf. Fig.~\ref{fig:EffectOfLapseTrick})
that choosing Eq.~(\ref{eq:SchwarzschildTime}) instead of 
$t_s=t$ significantly increases the accuracy of 
our extrapolation procedure during merger and ringdown.

Having computed the retarded time at each extraction radius,
we now consider the extracted waveforms as functions of retarded time
$u$ and extraction radius $\ra$, i.e. $\Psi_4^{22}(u,\ra)$.  At each
value of $u$, we have 
the phase and amplitude of
$\Psi_4^{22}$
at several extraction radii $\ra$. Therefore at each
value of $u$, we fit phase and amplitude separately to a polynomial in
$1/\ra$:
\begin{align}\label{eq:phi-Extrapolation}
  \phi(u,\ra)&=\phi_{(0)}(u)+\sum_{k=1}^n \frac{\phi_{(k)}(u) }{\ra^k},\\
\label{eq:A-Extrapolation}
  \ra\, A(u,r)&=A_{(0)}(u)+\sum_{k=1}^n \frac{A_{(k)}(u) }{\ra^k}.
\end{align}
The phase and amplitude of the desired asymptotic waveform are thus
given by the leading-order term of the appropriate polynomial, as a
function of retarded time:
\begin{align}
  \phi(u)&=\phi_{(0)}(u),\\
  \ra\, A(u)&=A_{(0)}(u).
\end{align}

\begin{figure}
  \includegraphics[width=\linewidth]{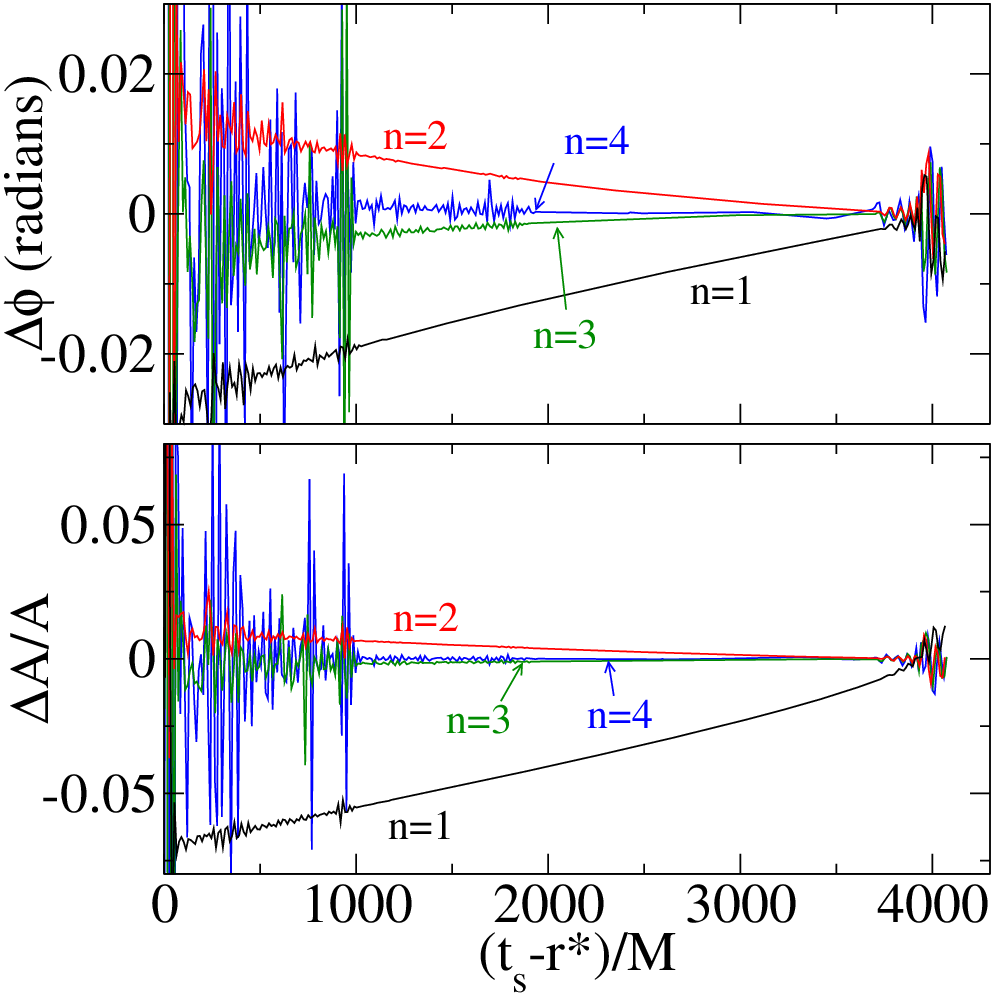}
  \caption{\label{fig:ExtrapolationConvergence} Convergence of
    extrapolation to infinity for extrapolation of order $n$.  For
    each $n$, plotted is the extrapolated waveform from run 30c1/N6
    using order $n+1$ minus the extrapolated waveform using order $n$.
    The top panel shows phase differences, the bottom panel shows
    amplitude differences.  No shifting in time or phase has been done
    for this comparison.  Increasing $n$ increases accuracy in smooth
    regions but also amplifies noise.}
\end{figure}

Figure~\ref{fig:ExtrapolationConvergence} shows phase and amplitude
differences between extrapolated waveforms that are computed using
different values of polynomial order $n$ in
Eqs.~(\ref{eq:phi-Extrapolation}) and~(\ref{eq:A-Extrapolation}).  For
the extrapolation we use waveforms extracted at radii $75M$, $85M$,
$100M$, $110M$, $130M$, $140M$, $150M$, $160M$, $170M$, $180M$,
$190M$, $200M$, $210M$, and $225M$.  From
Fig.~\ref{fig:ExtrapolationConvergence} it is clear that increasing
$n$ increases the accuracy of the extrapolation in smooth regions, but
also amplifies any noise present in the waveform.  Our preferred
choice, $n=3$, gives a phase error of $0.005$ radians and a
relative amplitude error of $0.003$ during
most of the inspiral,
and a phase error of $0.01$ radians and a relative amplitude error of
$0.01$ in the ringdown.  The junk radiation epoch
$t_s-r^\ast\lesssim 1000M$ has moderately larger errors than the ringdown.
If we were to choose instead $n=4$, we would gain higher accuracy in the
smooth regions at the expense of increased noise in the junk radiation epoch
and slightly larger errors
during the merger and ringdown.

\begin{figure}
  \includegraphics[width=\linewidth]{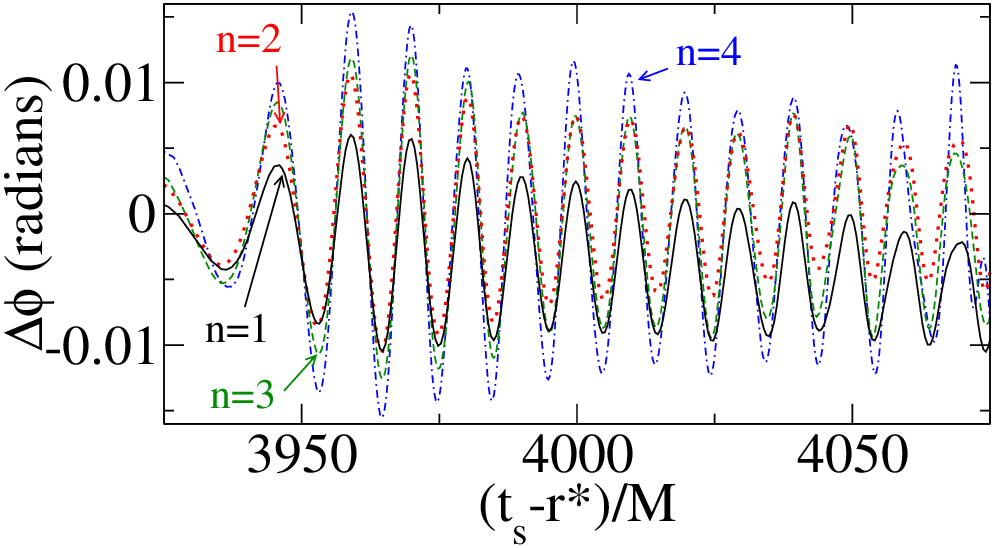}
  \caption{\label{fig:ExtrapConvergenceLateTimes} 
    Late-time phase
    convergence of extrapolation to infinity. Same as the top panel of
    Fig.~\ref{fig:ExtrapolationConvergence}, except zoomed to late
    times.  The peak amplitude of the waveform occurs at
    $t_s-r^\ast=3954M$.}
\end{figure}

\begin{figure}
  \includegraphics[width=\linewidth]{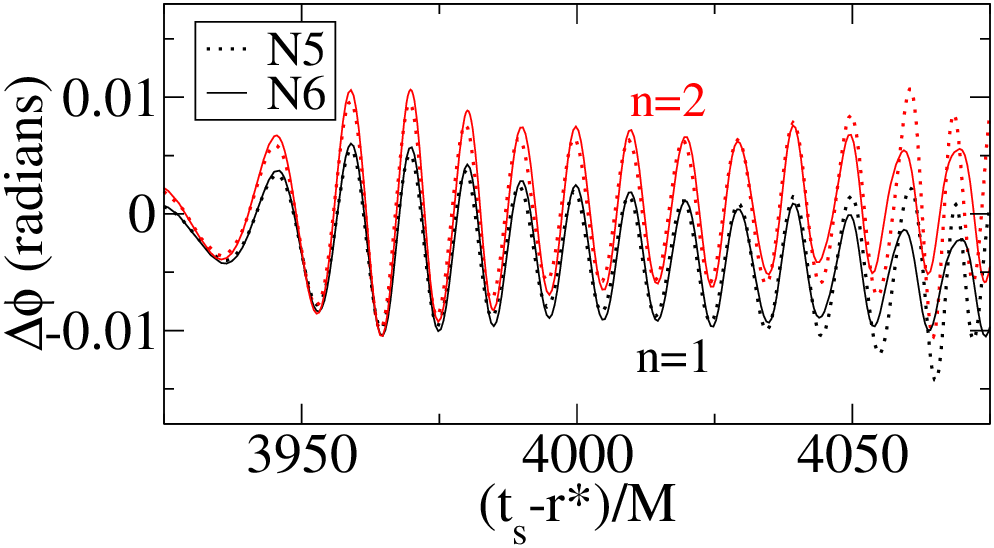}
  \caption{\label{fig:ExtrapConvergenceLev5vsLev6}
    Effect of numerical resolution on extrapolation to
    infinity.  The solid
    curves are identical to the ``n=1'' and ``n=2'' curves from
    Fig.~\ref{fig:ExtrapConvergenceLateTimes}.  The dotted
    curves are the same quantities computed using the lower
    resolution run 30c1/N5.}
\end{figure}

Figure~\ref{fig:ExtrapConvergenceLateTimes} is the same as the
top panel of Fig.~\ref{fig:ExtrapolationConvergence}, except zoomed to
late times.  Note that during merger and ringdown,
the extrapolation procedure does not converge with
increasing extrapolation order $n$: the phase differences are slightly
larger for larger $n$.  This lack of convergence suggests that the
nonextrapolated numerical waveform contains some small
contamination that does not
obey the fitting formulae, Eqs.~(\ref{eq:phi-Extrapolation})
and~(\ref{eq:A-Extrapolation}).  
Figure~\ref{fig:ExtrapConvergenceLev5vsLev6} shows the n=1 and
  n=2 convergence curves from 
  Fig.~\ref{fig:ExtrapConvergenceLateTimes}, but computed for two different
  numerical resolutions, 30c1/N5 and 30c1/N6.  The N5 and N6 lines are very
  close to each other in this figure, indicating that the lack of
  convergence with extrapolation order $n$ is not dominated by
  insufficient numerical resolution. 
We suspect that the main contribution is instead
due to gauge effects.
Such gauge effects
might be reduced by improving the gauge conditions in the
numerical simulation or by adopting more sophisticated wave extraction
and extrapolation algorithms that better compensate for dynamically varying
gauge fields.

\begin{figure}
  \includegraphics[width=\linewidth]{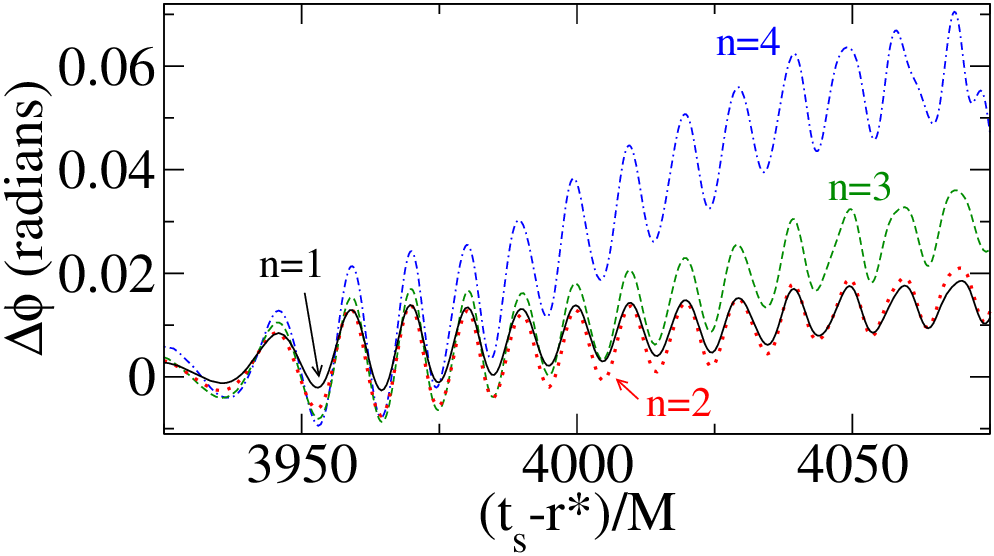}
  \caption{\label{fig:EffectOfLapseTrick} 
    Effect of $t_s$ on extrapolation to infinity.
    Same as Fig.~\ref{fig:ExtrapConvergenceLateTimes}, except the
    quantity $t_s$ that appears 
    in the retarded time, Eq.~(\ref{eq:RetardedTime}),
    is chosen to be coordinate time $t$ rather than 
    the integral in Eq.~(\ref{eq:SchwarzschildTime}).
    Note the difference in vertical scale between this figure
    and Fig.~\ref{fig:ExtrapConvergenceLateTimes}.}
\end{figure}

Indeed, we have already made a first attempt at correcting
for a time-dependent
lapse function by using $t_s$ from Eq.~(\ref{eq:SchwarzschildTime}) to
compute the retarded time.  
Figure~\ref{fig:EffectOfLapseTrick} illustrates the importance of this
correction.  Figures~\ref{fig:ExtrapConvergenceLateTimes}
and~\ref{fig:EffectOfLapseTrick} differ
only in the choice of $t_s$ used to compute the retarded time: In 
Fig.~\ref{fig:ExtrapConvergenceLateTimes}, $t_s$ is
obtained from Eq.~(\ref{eq:SchwarzschildTime}),
and in Fig.~\ref{fig:EffectOfLapseTrick}, 
$t_s$ is simply the coordinate time $t$.
Using the naive choice $t_s=t$ clearly results in much larger phase differences
that diverge with increasing $n$ and grow in time.

\begin{figure}
  \includegraphics[width=\linewidth]{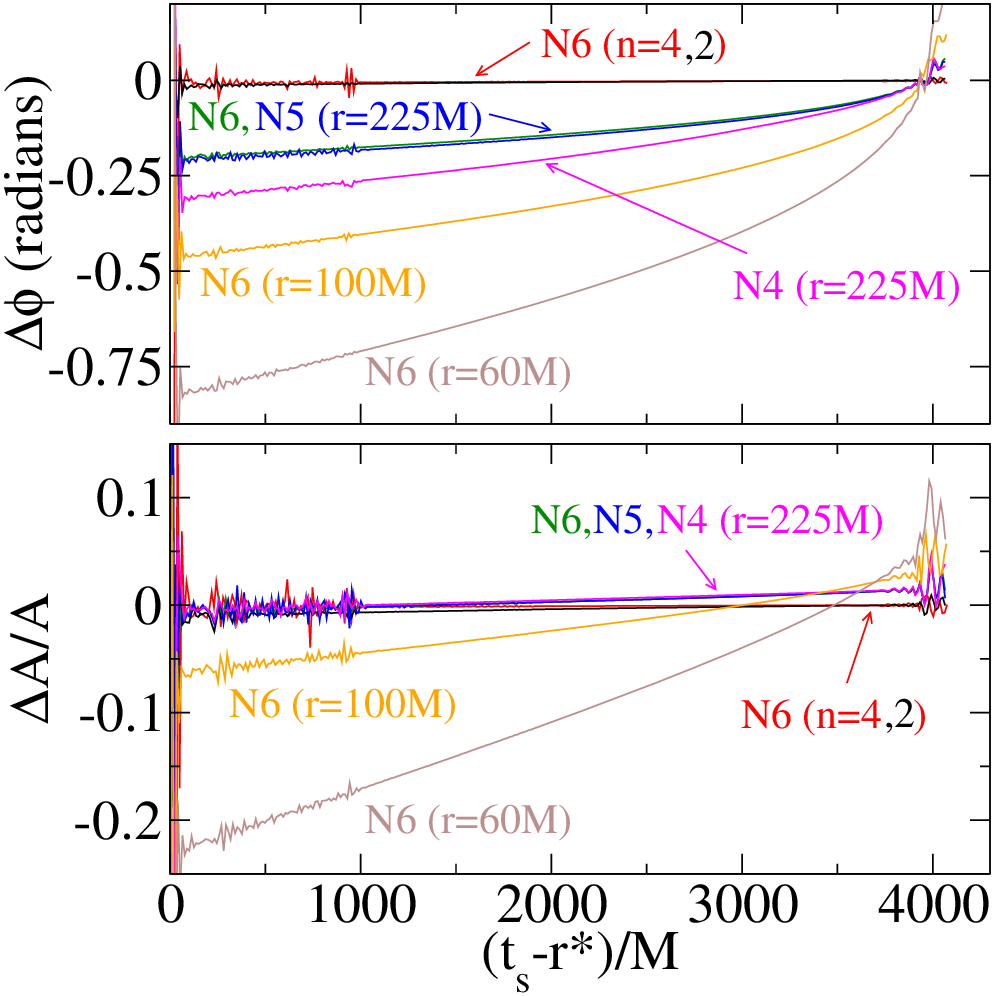}
  \caption{\label{fig:ExtrapolationVsExtraction} Comparison of
    extrapolated and nonextrapolated waveforms.  Plotted are
    differences between selected waveforms and the 30c1/N6 waveform
    extrapolated to infinity using $n=3$.  Each selected waveform is
    labeled by the numerical resolution (N4, N5, or N6), and either
    the extraction radius (for nonextrapolated waveforms) or the
    extrapolation order (for extrapolated waveforms).  Each waveform
    has been shifted in time and phase so as to minimize the
    least-squares difference from the N6, $n=3$ waveform.  The top
    panel shows phase differences, the bottom panel shows amplitude
    differences. Differences between extrapolated and
    nonextrapolated waveforms are much larger than differences between
    different extrapolation orders.  Phase differences between resolutions
    N5 and N6, and amplitude differences between all three resolutions,
    are indistinguishable on the plot.
  }
\end{figure}

In Fig.~\ref{fig:ExtrapolationVsExtraction} we examine the difference
between extrapolated waveforms and waveforms that have been extracted
at a finite radius.  We compare our preferred waveform, 30c1/N6
extrapolated to infinity using $n=3$, versus nonextrapolated
waveforms and versus extrapolated waveforms with different values of
$n$.  Because the extrapolated and nonextrapolated waveforms differ
by overall time and phase offsets which are irrelevant for many
purposes, each waveform in Fig.~\ref{fig:ExtrapolationVsExtraction}
has been shifted in time and phase so as to best match with the $n=3$
extrapolated waveform.  This best match is determined by a simple
least-squares procedure: we minimize the function
\begin{equation}
  f(t_0,\phi_0)\!=\!\sum_i\!\!\left(\!A_1(t_i)e^{i\phi_1(t_i)}\!-\!%
    A_2(t_i+t_0)e^{i(\phi_2(t_i+t_0)+\phi_0)}\right)^2\!\!\!,
\end{equation}
by varying $t_0$ and $\phi_0$.  Here $A_1$, $\phi_1$, $A_2$, and
$\phi_2$ are the amplitudes and phases of the two waveforms being
matched, and the sum goes over all times $t_i$ at which waveform 1 is
sampled.

We find from Fig.~\ref{fig:ExtrapolationVsExtraction} that
extrapolation to infinity has a large effect on the phase of the final
waveform and a much smaller effect on the amplitude, when comparing to
data extracted at our outermost extraction radius, $r=225M$.  The
$r=225M$ waveforms have an accumulated phase difference of $0.2$
radians relative to the extrapolated waveform, much larger than the
difference between different extrapolation orders or different
numerical resolutions.  For extraction at smaller radii, the
differences are larger still, the $r=60M$ waveform having a phase
difference of $0.8$ radians and amplitude difference of $20$ percent
compared to the extrapolated waveform.  We find that the phase
  differences between extrapolated and nonextrapolated waveforms
  scale quite accuratly like $1/r$, and the amplitude differences
  scale roughly like $1/r^{2.5}$, where $r$ is the extraction radius.
  These scalings seem to be related to near-field effects, for which
  one expects scalings like $1/r$ in phase and $1/r^{2}$ in
  amplitude~\cite{Boyle2008}.

\begin{figure*}
\includegraphics[width=\textwidth]{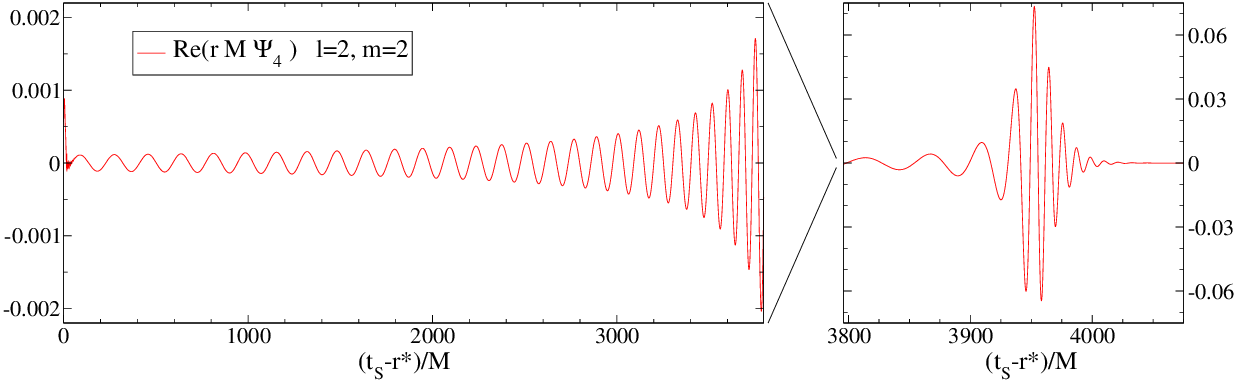}\\[1em]

\includegraphics[width=\textwidth]{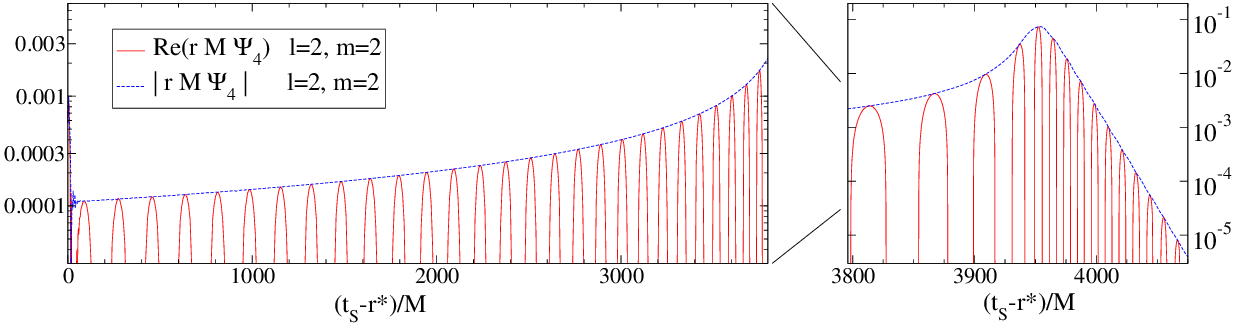}
\caption{\label{fig:ExtrapolatedWaveform}Final waveform, extrapolated to infinity.  The top panels
  show the real part of $\Psi_4^{22}$ with a linear y-axis, the bottom
  panels with a logarithmic y-axis.  The right panels show an
  enlargement of merger and ringdown.  }
\end{figure*}

Figure~\ref{fig:ExtrapolatedWaveform} presents the final
waveform after extrapolation to infinite radius.  There are 33
gravitational-wave cycles before the maximum of $|\Psi_4|$.  The
simulation is further able to resolve 10 gravitational-wave cycles
during ringdown, during which the amplitude $|\Psi_4|$ drops by 
four orders of magnitude.

\section{Discussion}
\label{sec:Discussion}

We have presented the first spectral computation of a binary black
hole inspiral, merger, and ringdown, and we have extracted accurate
gravitational waveforms from our simulation. A key ingredient in
handling the merger phase is a choice of gauge that causes the
individual holes to expand in coordinate size. This eliminates the
coordinate singularities that prevented our earlier simulations from
continuing through merger.  The largest downside to the gauge used
here is that the success of the method depends sensitively on some of
the gauge parameters, namely $\sigma_1$ and $\sigma_2$ in
Eq.~(\ref{eq:GaugeRolloff}), and $\xi_1$ and $\xi_3$ in
Eqs.~(\ref{eq:Hevolutiont}) and~(\ref{eq:Hevolutioni}).  If these
parameters are chosen poorly, the characteristic fields at the
excision boundaries fail to be purely outgoing (i.e. into the holes)
at some instant in time, causing the code to terminate due to lack of a
proper boundary condition at an excision boundary.  An alternative
approach to gauge conditions for the generalized harmonic
system~\cite{Lindblom2007} is in progress, and promises to be more
robust.

We compute the spin of the final black hole with three distinct
diagnostics, one based on approximate rotational Killing vectors, the
others based on the minimum and maximum of the scalar curvature of the
apparent horizon ($\chi_{\rm AKV}$, $\chi_{\rm SC}^{\rm min}$, and
$\chi_{\rm SC}^{\rm max}$ in the language of Appendices A and B
of~\cite{Lovelace2008}).  We find that these diagnostics agree to an
exquisite degree.  Since these diagnostics coincide exactly for a Kerr
black hole, this suggests that the final state is indeed a Kerr
black hole.  The uncertainty of the final spin quoted in
Sec.~\ref{sec:PropertiesOfSolution} is due to numerical truncation
error, (i.e. differences between resolutions 30c1/N5 and 30c1/N6),
rather than due to differences between spin diagnostics, and we
find $S_f/M_f^2 =0.68646\pm 0.00004$, and $M_f=(0.95162\pm 0.00002)M$.

The physical waveform at infinity produced by any numerical relativity
code should of course be independent of the coordinates used during
the simulation.  However, in practice it is difficult to remove
coordinate effects from the waveform for several reasons. First,
waveforms are typically extracted on coordinate spheres (not geometric
spheres) of finite radius as functions of coordinate time (which may not
agree with proper time at infinity). 
Second, the extracted waveform on a given sphere is
typically expanded in spin-weighted spherical harmonics ${}_sY_{\ell
  m}(\theta,\phi)$ using the $\theta$ and $\phi$ coordinates from the
simulation rather than some geometrically defined $\theta$ and $\phi$
coordinates.  Finally, standard formulae equating $\Psi_4$ with the
asymptotic radiation field assume that $\Psi_4$ is computed at
infinity.  Such gauge ambiguities can be significant for the
accuracy of waveforms from numerical
simulations~\cite{Nerozzi2006,Pazos2006,Lehner2007}.  Indeed, if we
choose a deliberately ``bad'' gauge just after merger by omitting the
factor $e^{-r''^2/\sigma_3^2}$ in the function $f(x,t)$
[cf. Equation~(\ref{eq:GaugeRolloff2})], we find that the lapse function
oscillates in time even at large distances, and that the resulting
waveform extracted at a finite radius differs
by more than a radian in
phase from the waveform presented here.
We defer further discussion of gauge effects on the waveform to
a future paper.

We have also shown that extrapolation of waveforms to infinity is
crucial: waveforms extracted at a finite radius differ (particularly
in phase) from waveforms extrapolated to infinity by far more than the
numerical errors, as shown in
Fig.~\ref{fig:ExtrapolationVsExtraction}.  Although it is likely that
the need for extrapolation may be somewhat reduced by more
sophisticated algorithms for wave extraction at finite radius, it
appears that most of the difference between waveforms that have and
have not been extrapolated to infinity is due to physics (in the form
of near-zone effects) rather than to gauge and tetrad
ambiguities~\cite{Boyle2008}.

We are currently extending our methods to binary black holes with
unequal masses and nontrivial spins.  Inspiral simulations for these
more generic systems have already been computed by our code; it
remains to be seen whether mergers of more generic black hole systems
can be simulated with the methods described here, or whether
alternative gauge conditions, such as those described in
Ref.~\cite{Lindblom2007}, will be necessary.

It would be interesting to compare the waveforms presented here with
those from other groups computing binary black hole mergers,
particularly since other groups use different numerical methods,
different formulations of the equations, and different gauge
conditions than our group.  Several such comparisons are presently
under way.

Waveforms are available at \\ http://www.black-holes.org/Waveforms.html.

\begin{acknowledgments}

We thank Luisa Buchman,
Luis Lehner, Frans Pretorius, Bela Szil\'agyi, Saul
Teukolsky, and Kip Thorne for helpful comments, Fan Zhang
for help with the extrapolation code, 
and Geoffrey Lovelace and Rob Owen for 
providing the diagnostics used to measure the final spin.  
We are especially grateful to Lee
Lindblom for numerous suggestions, ideas, and discussions that
significantly contributed to the success of the methods described
here.  This work was supported in part by grants from the Sherman
Fairchild Foundation to Caltech and Cornell, and from the Brinson
Foundation to Caltech; by NSF grants PHY-0601459, PHY-0652995,
DMS-0553302 and NASA grant NNG05GG52G at Caltech; by NSF grants
PHY-0652952, DMS-0553677, PHY-0652929, and NASA grant NNG05GG51G at
Cornell.  We thank NASA/JPL for providing computing facilities that
contributed to this work. 
Some of the simulations discussed here were produced with LIGO
Laboratory computing facilities. LIGO was constructed by the
California Institute of Technology and Massachusetts Institute of
Technology with funding from the National Science Foundation and
operates under cooperative agreement PHY-0107417.

\end{acknowledgments}



\end{document}